# TMS–EEG Reliability:

# Bridging the Gap to Clinical Use

Giacomo Bertazzoli[1,2,3,4,5]*, Carlo Miniussi[2], Petro Julkunen[3,4], Marta Bortoletto[1]

1 Neurophysiology Lab, IRCCS Istituto Centro San Giovanni di Dio Fatebenefratelli, Brescia, Italy[#]

2 Center for Mind/Brain Sciences CIMeC, University of Trento, Rovereto, Italy[#]

3 Department of Neurology, Brigham and Women's Hospital, Boston, MA, United States[@]

4 Department of Neurology, Massachusetts General Hospital, Boston, MA, United States[@]

5 Harvard Medical School, Cambridge, MA, United States[@]

6 Department of Clinical Neurophysiology, Kuopio University Hospital, Kuopio, Finland

7 Department of Technical Physics, University of Eastern Finland, Kuopio, Finland

*Corresponding author:

Giacomo Bertazzoli, PhD

IRCCS Istituto Centro San Giovanni di Dio Fatebenefratelli, via Pilastroni 4, 25125 Brescia, Italy.

E-mail: giacomo.bertazzoli@cognitiveneuroscience.it

[#] Study performed with these affiliations

[@] Current affiliations

# Highlights

- TMS-EEG-derived indices are appealing biomarkers in several neurological conditions.
- However, we argue that reliability assessment of these measures has not met standards for clinical practice
- We describe reliability assessments needed to gauge the clinical utility of TMS-EEG-derived indices

# Abstract

Concurrent transcranial magnetic stimulation (TMS) and electroencephalography (EEG), i.e., TMS-EEG, may expand the potential clinical applications of TMS beyond the conventional evaluation of the cortico-spinal tract and motor cortices. TMS-evoked potentials (TEPs) have been used in clinical research to assess cortical excitability and effective connectivity between cortical areas in several psychiatric and neurological disorders, and are evaluated as an appealing candidate biomarker to help diagnosis and/or prognosis. However, while TMS is a well-established diagnostic tool, TMS-EEG has not yet met standards necessary for clinical implementation. Considering any evidence-based clinical applications of TMS-EEG, a crucial point is that reliability assessments of TEPs are often missing. Here, we describe reliability assessments needed to gauge the clinical utility of TEPs. Specifically, we review current literature on reliability and describe multiple theoretical and statistical elements that are included within this term. Then, we present current knowledge on TEP reliability and highlight the elements of reliability that need to be implemented to enable a unified evidence-based reliability assessment of TMS–EEG as a clinical tool.

## Keywords

## 1. Introduction

Transcranial magnetic stimulation (TMS) can be combined with neurophysiological and neuroimaging techniques to measure the state of the nervous system.

The most common combination is with electromyography (EMG). In this case, TMS allows to measure motor-evoked potentials (MEPs) for indirectly mapping the motor cortex representation of muscles and investigating the excitability, conduction, and integrity within or between motor cortices as well as the cortico-spinal tract (Barker, Jalinous, and Freeston 1985; Chen et al. 2008a; Kobayashi and Pascual-Leone 2003; Rossini et al. 2015). TMS-induced MEPs are now an established measure applied in clinical practice to examine the functional state of the corticospinal pathways in several diseases involving motor dysfunction (Groppa et al. 2012).

More recently, TMS has been combined with neuroimaging techniques, such as electroencephalography (EEG), with the final aim to extend TMS-derived biomarkers to a wider range of psychiatric and neurological disorders. TMS-EEG coregistration allows for assessing cortical excitability and function more directly from the cortex (Ilmoniemi and Kičić 2010; Siebner et al. 2009). Specifically, the TMS-evoked cortical response measured with EEG, i.e., TMS-evoked potentials (TEPs), provide information on cortical excitability, i.e., the strength at which the cortex responds to the TMS (Bonato, Miniussi, and Rossini 2006; Casula et al. 2022; Kähkönen et al. 2004), and effective connectivity, i.e., the spread of activation through structural and functional connections (Bortoletto et al. 2015, 2021; Momi, Ozdemir, Tadayon, Boucher,

Di Domenico, et al. 2021; Ozdemir et al. 2020; Zazio et al. 2022). Recent studies have shown the potential application of TEPs to measure alterations in cortical activity for several psychiatric conditions, such as major depressive disorder, bipolar disorder and schizophrenia, (Canali et al. 2015; D'Agati et al. 2014; Kirkovski et al. 2016; Levit-Binnun et al. 2009; Naim-Feil et al. 2016; Noda, Barr, et al. 2018; Noda, Zomorrodi, et al. 2018), neurological conditions, such as Alzheimer's disease (Bagattini et al. 2019; Casarotto et al. 2011; Ferreri et al. 2016, 2021; Julkunen et al. 2008; Koch et al. 2018; Kumar et al. 2017), disorders of consciousness (among which Arai et al. 2021; Bai et al. 2016; Bodart et al. 2017; Casarotto et al. 2016; Ferrarelli et al. 2010; Formaggio et al. 2016; Gosseries et al. 2014; Massimini et al. 2012; Ragazzoni et al. 2013; Rosanova et al. 2009, 2012; Sarasso et al. 2014 for review see Ragazzoni et al. 2017) and stroke (Bodart et al. 2017; Borich et al. 2016; Cipollari et al. 2015; Manganotti et al. 2015; Pellicciari et al. 2018).

However, despite the perspectives of its clinical application, TMS–EEG is still far from achieving clinical standards. The lack of systematic quantification of "accuracy and reliability" needed for planning clinical trials is preventing TMS-EEG from being used in clinical contexts (Julkunen et al. 2022). As detailed more deeply in the next section, reliability assesses the ability of a measure to remain stable in unchanging individuals. This feature is key for diagnostic and prognostic biomarkers as it allows to disentangle changes due to real physiological events from noise. Therefore, proving that TMS-EEG-based measures possess high reliability is a fundamental step toward clinical applications.

The reliability of the TMS–EEG signal has been investigated previously, highlighting that some features of TEPs are stable within-subjects at various time intervals, ranging from days to months (Table 1). However, the current reliability assessments of TMS–EEG-based measures are not sufficient to determine clinical utility of TMS-EEG. Specifically, reliability assessments of TMS–EEG has either failed to test reliability in a strict sense or has focused on "relative" reliability - a particular kind of reliability that depends on the population studied rather than the measure itself. Focusing on relative reliability leads to poor generalizability, as we further elaborate below.

Relying on the current partial assessment of TMS-EEG-based measures' reliability, incorrect conclusions could be drawn affecting the estimates of clinical applicability of TMS-EEG. Other TMS (Beaulieu et al. 2017; Schambra et al. 2015) and non-TMS (Weir 2005) related measures have faced these same kinds of issues in the past, and we argue here that the same solutions to reliability issues facing other fields will be beneficial to the TMS-EEG field.

## 2. Consensus terminology and definition of reliability

When discussing the reliability of a biomarker, or of biological measures more generally, there is often inconsistency about what reliability really is due to the use of different terminology to convey overlapping meaning, for example, reproducibility, validity, stability, and consistency. Moreover, some of these terms relate to different statistical concepts, leading to further misinterpretations (Mokkink et al. 2010; de Vet et al. 2006).

Adopting consensus terminology from Mokkink et al., (2010), the usefulness of a neurophysiological measure as a clinical or diagnostic tool depends on its *validity*, *responsiveness*, and *reliability*. These domains, in the context of neurophysiology, cover the ability of a measure to assess a target neurophysiological process (validity), detect neurophysiological changes when they occur (responsiveness), and remain stable in unchanging conditions (reliability) (see Table 2 in Mokkink et al. 2010).

This conception of reliability is operationalized by breaking it down into two equally important kinds of reliability, relative and absolute, which refer to strictly different concepts and operationalizations (Atkinson and Nevill 1998; Beaulieu, Massé-Alarie, et al. 2017; McManus 2012; Schambra et al. 2015; Weir 2005).

*Relative reliability* refers to the degree to which unchanging individuals maintain their position relative to each other across repeated measures (Streiner and Norman 2016; Terwee et al. 2007; de Vet et al. 2006; Weir 2005). In other words, if we take the same measure at two time points (e.g., T0 and T1) in the same cohort of individuals, relative reliability refers to the degree to which the ranking of individuals based on the measure is the same or similar between the two time points. Therefore, assessing relative reliability allows us to test whether a measure has enough between-subject variance, in other words that the measure is "heterogenous enough" between subjects, to tell individuals apart stably over time. This kind of reliability can be assessed with, for example, the intraclass correlation coefficient (ICC)

$$Relative\ Reliability = ICC = \frac{between-subject\ variance}{between-subject\ variance + residual\ error} = \frac{\sigma_s^2}{\sigma_s^2 + \sigma_e^2}, \quad (1)$$

where the between-subject variance and the residual error are estimated using mean square values from an analysis of the variance (ANOVA). Based on the design of the experiment and the desired reliability outcome, the terms of the above equation can be estimated in different ways, which is beyond the scope of this review (see Shrout and Fleiss 1979; Weir 2005). The closer the ICC value is to 1, the higher the relative reliability. Values near 0 indicate poor to null relative reliability, and negative values are not theoretically meaningful, but possible in some cases (Giraudeau 1996). When interpreting ICC values, many studies refer to Shrout (1998), who defined ranges of relative reliability with 0.00 – 0.10 as "virtually none", 0.11 – 0.40 as "slight", 0.41 – 0.60 as "fair", 0.61 – 0.80 as "moderate" and 0.81 – 1.00 as "substantial".

*Absolute reliability,* on the other hand, captures the degree to which repeated measures, taken for the same unchanged individual at different time points (e.g., T0 and T1) remain the same. Absolute reliability can thus be considered the absolute difference between a measure taken at two different time points. If the absolute difference is small, the absolute reliability is high, while if the difference is high, then absolute reliability is low. Absolute reliability is often operationalized with the standard error of a measurement, which is abbreviated SEM, but to avoid confusion with the standard error of the mean, the abbreviation *SEMeas* was adopted here as in Schambra et al., 2015 (Beaulieu et al. 2017; Mokkink et al. 2010; Terwee et al. 2007; de Vet et al. 2006). The *SEMeas* typically captures the residual error (but in one of its variants, it can capture both residual and systematic errors) of a repeated measure (Hopkins 2000). The smaller the *SEMeas,* the lower the measurement error across sessions and the more consistent the measure is:

$$Absolute\ reliability = SEMeas = \sqrt{residual\ error} = \sqrt{\sigma_e^2} \qquad (2)$$

An important feature that differentiates the SEMeas from the ICC is that the SEMeas does not depend on the variability of the population in exam, it is not influenced by between-subject variability. Rather, it captures the "typical error" of a measure, which mostly depends on the technique and is independent from the population (Hopkins 2000). *SEMeas* is often used to derive a more immediate index, the *smallest detectable change* (*SDC*, also called *minimal difference*) (Beaulieu et al. 2017; Schambra et al. 2015; Terwee et al. 2007; Weir 2005), which indicates the smallest change needed in a measure that can be considered a "true" change, for example, the smallest change in µV of a TEP component in a test-retest paradigm to consider it a real effect of the manipulation rather than random fluctuation.

Beside the ICC and SEMeas/SDC, other indices can be computed to assess relative and absolute reliability. Worth mentioning here is the concordance correlation coefficient (CCC), which has often been used in TMS-EEG literature. The CCC was created by Lin (1989) to assess absolute reliability, but subsequent studies have demonstrated that the CCC still depends on sample variability (Barnhart et al. 2007) and, in some cases, is equivalent to ICC (Carrasco and Jover 2003).

### 2.1 Assessing relative or absolute reliability: an example of the relation between the ICC and SEMeas/SDC

Suppose we want to assess the reliability of one TEP component's latency in the same individuals across multiple time points, we would collect TEPs at two time points (T0 and T1), e.g., one month apart, and calculate the latency of the component. Four sets of fictional data are provided in Table 2, in which, per each set, we depict the latency of a TEP component in eight subjects and their analysis of the variance (used to compute the ICC and the SEMeas/SDC; example inspired by Weir, 2005).

In panel A, the ICC between the two session is 'substantial' (0.85) according to Shrout (1998), which suggests that the latency this component has high relative reliability. However, the *SEMeas* and *SDC* are also high and thus suboptimal (4.42 ms and 12.25 ms, respectively), which means that the change in the latency of the component after an intervention could be considered a "true" change only if it exceeds 12.25 ms because the TMS-EEG signal is noisy. Absolute reliability is thus low.

We might then ask, why do the relative and absolute reliability estimation conflict? One explanation for the discrepancy between relative and absolute reliability is visualized in Figure 1A. ICC is high is because the relative position or ranking of the subjects between T0 and T1 remained the same, at least for most subjects. Since ICC gauges the ability of an index to rank individuals in the same way across repeated measures, the ICC remains high despite the change in the absolute value of the measure between T0 and T1 is high. Note also that the ICC is the proportion of the between-subject variance to the total variance, which, as reported in the ANOVA (panel A in Table 2), favors the between-subject variance. Such a high between-subject variance could be caused by the technique capturing a phenomenon that varies drastically between subjects or because the specific population examined has high variance. For example, patients in different stages/states of a disease could vary greatly in their neurophysiology. Conversely, the SEMeas and the SDC are sensitive only to the change in absolute value within subjects and between time points rather than the relative ranking of individuals between the two time points. Consequently, despite the TEP component latency showing a high *ICC*, the high *SEMeas*/*SDC* indicate that this measure is not stable within subjects across time, and is thus suboptimal if we want to detect small, but meaningful changes within subjects when they occur.

Table 2 (panel C) and Figure 1C describes an example similar to the one described above. The only difference is represented by the between-subject variance that, in this case, is reduced. Note that the differences in scores within-subjects are identical to those in panel A (Table 2). However, the shrinking of the between-subject variance caused the *ICC* to decrease to 0.35, which would be described as 'virtually none' by Shrout (1998), while the *SEMeas* and *SDC* remain the same. This highlights the population-dependence of the ICC: when using the same measure i.e., the TEP component latency, just in another population with its own between-subject variance, the ICC varies. Instead, the *SEMeas/SDC* remain the same because of their dependency on the variability of the technique instead of the population. This discrepancy in ICC is meaningful when, for example, inferring reliability of measure in a patient population from estimations made in healthy volunteers or young individuals.

As another example, panel B of Table 2 depicts a scenario in which the between-subject variability is similar to that found in panel A, but the difference in measure scores between the two sessions is reduced. In this case, the high between-subject variance yields a high *ICC* of 0.98, and the low residual error yields a low (optimal) *SEMeas*/*SDC* of 1.30 ms and 3.60 ms, respectively. This is visualized in Figure 1B, showing how the relative ranking of the subjects does not change (high *ICC*) while the change between time points within-subjects remains low (low *SEMeas/SDC*).

Finally, we could face a scenario like what is depicted in panel D of Table 2, which is similar to panel B, but with reduced between-subject variance. In this case, the lower between-subject variance prevents the relative ranking of the subjects from remaining stable across repeated measures, yielding a low *ICC* of 0.49. However, the low residual error between measures within-subjects keeps a low (optimal) *SEMeas/SDC* of 1.41 ms and 3.92 ms, respectively (Figure 1D).

To conclude, the *ICC* can only be interpreted as a reflection of the examined population, as there are cases in which the *ICC* is low (suboptimal) even when the absolute differences between repeated measures are low (Table 2 panel D) or cases in which the *ICC* is high even when the differences between repeated measures are high (Table 2 panel A). On the other hand, the SEMeas/SDC indexes how reliable a measure remains across populations as long as the same technique is employed (Beaulieu et al. 2017; Schambra et al. 2015; Weir 2005). Therefore, complementing the ICC with the SEMeas/SDC i.e., complementing relative reliability with absolute reliability estimates, provides a more complete picture of a measure's reliability. This is useful not only for determining clinical potential of measure, in this case TMS-EEG, but also for avoiding erroneous conclusions made based on only one of the two reliability estimates.

A feature that distinguishes measures of relative reliability, such as the ICC, with measures of absolute reliability, here the SEMeas/SDC, it is also their interpretation ease. A high or low ICC is fairly easy to define, with 1 reflecting perfect relative reliability and 0 reflecting no relative reliability. In addition, pre-defined interpretations and scales, like from Shrout (1998), can be used to standardize ICC interpretations and interpret intermediate values. On the contrary, the interpretation of absolute reliability indices like the SEMeas/SDC is less straightforward. For instance, whether an SEMeas/SDC value should be considered high (suboptimal) or low (optimal) depends entirely on the technique employed and the phenomenon studied. So, while interpretations of SEMeas/SDC are possible to standardize for each technique, comparing the absolute reliability assessed by SEMeas/SDC between techniques, for example when deciding which technique to use in a clinical context, remains a challenge.

*2.2. Relative and absolute reliability in clinical applications: diagnostic and prognostic biomarkers*

As discussed by Schambra et al. (2015) and Beaulieu et al., (2017) for MEPs, the impact of between-subject variance on the ICC, described as a feature above, could be interpreted as a limitation of ICC. However, the ICC, when properly interpreted as an assessment of relative reliability, constitutes a fundamental step in establishing of novel diagnostic biomarker for a population, which are used to assess the presence of a pathology. As a diagnostic biomarker, we refer to a measure that can accurately discriminate between the presence or absence of a pathology. Assuming that the validity (ability of a measure to assess neurophysiological processes) and responsiveness (ability of a measure to detect neurophysiological changes when they occur) of a measure are established, the measure would still need to distinguish individuals in a stable fashion over time, which is relative reliability. In other words, the relative ranking of unchanging individuals should be the same across time. If a measure does not have high relative reliability, diagnoses would be unstable. Without adequate relative reliability, the relative ranking of patients might change at every measurement, making it difficult to determine whether an individual qualifies as having the pathology or not, even if absolute reliability is high (for example, in Figure 1D). Therefore, high between-subject variance might be necessary to establish a diagnostic biomarker with high relative reliability. Even if the error between measurement of the same subject is high (low absolute reliability), if each individual preserves their position relative to the others, the measure can be considered a reliable diagnostic biomarker in a relative sense (Figure 1A).

For the establishment of a novel prognostic biomarker, which is used to assess or make inferences about the progression of a pathology, high absolute reliability becomes the priority. As a prognostic biomarker, we refer to a measure/index that can accurately track certain biological phenomena within individuals. Again, if the measure in the exam has already proven valid and responsive, such a measure would then need to be able to capture meaningful, but sometimes subtle, changes in the examined phenomenon to track of its evolution over time. The minimum amount of change that a prognostic biomarker should detect depends on the measured phenomenon itself, but, in general, the smaller this minimum change, the more sensitive the biomarker. Therefore, if a measure has high ICC (relative reliability) and high (suboptimal) SEMeas/SDC, it would be difficult to track small changes that could be meaningful, since the intrinsic error of the technique is high (see Figure 1A). In an opposite scenario, low ICC, but low (optimal) SEMeas/SDC could justify the use of the measure as a prognostic biomarker of the studied phenomenon (see, for example, Figure 1D).

In combination, a measure that possesses both high absolute reliability and high relative reliability could be used as both a prognostic and diagnostic biomarker (see Figure 1B). One such measure would be sensitive enough to detect small, meaningful changes (high absolute reliability and prognostic utility), while also discriminating between the presence or absence of a pathology through stable relative rankings (high relative reliability and diagnostic utility).

In the next sections, we describe the current state of reliability assessments for TMS-EEG-based measures. We describe both studies that assess reliability in a strict sense, relative or absolute reliability, and studies that assessed reliability through other means (e.g., *t* tests, correlations, etc.)

## 3. Previous assessments on the reliability of TMS–EEG-based measures

The reliability of TMS–EEG-based measures has been explored previously by examining the TEP wave directly, but also by examining TEP-derived indices, such as long-interval cortical inhibition, interhemispheric signal propagation, and interhemispheric signal balance (see Table 1). In the first case, two main approaches have been taken: extracting TEP peak amplitude and latency or testing the continuous wave point-by-point.

Repeated measures similarities and differences in TEP peaks were first assessed by Lioumis et al. (2009), who stimulated the primary motor cortex (M1) and DLPFC at various intensities in two separate sessions one week apart. Pearson's *R* showed high linear correlation between amplitudes and latencies of peaks across time points, and *t*-tests showed no significant differences between most of the peaks across time points.

Similarly, Kerwin et al. (2018) tested the relative and absolute reliability of DLPFC TEP peaks between different time intervals and trial combinations with the concordance correlation coefficient (CCC) (Carrasco and Jover 2003; King et al. 2007; Lin 1989) and SDC, respectively. They reported high relative and absolute reliability of late peaks N100 and P200, while mixed results were found for early peaks. Recently, de Goede et al (2020) explored the relative reliability of M1 TEPs by means of ICC, finding poor (0.37 – 0.49) to moderate (0.60 – 0.75) relative reliability for components N100 and P180, depending on the stimulation protocol (single-pulse vs paired-pulse TMS).

Finally, Bertazzoli et al. (2021) tested the relative reliability of DLPFC and inferior parietal lobe (IPL) TEP peak amplitudes and latencies with the CCC. The results were in line with previous work, with increasing relative reliability from early to late peaks for both areas.

Assessing point-by-point reliability of whole-epoch TEP responses has been problematic, given that established indices of reliability would need to be calculated on big data matrices. Therefore, other strategies have been employed. Some studies investigated the differences in the whole TEP response between repeated sessions using cluster-based permutation *t*-tests, reporting no consistent significant differences in TEPs across time points (Bertazzoli et al. 2021; ter Braack et al. 2019; de Goede et al. 2020; Mancuso et al. 2021). Other studies investigated the similarity of the TMS–EEG response across time points by testing the correlation of the response in time and/or space dimensions. In other words, running correlations tests at each point in time across electrodes between sessions (spatial correlation) or testing correlation at each electrode in time between sessions using the whole signal or dividing the signal into time chunks (temporal correlation) (Bertazzoli et al. 2021; Conde et al. 2019; Ozdemir et al. 2020). These studies report stronger correlations from early to late latencies, highlighting the stability of late components compared to early components (Bertazzoli et al. 2021; Kerwin et al. 2018; Momi et al. 2021; Ozdemir et al. 2020).

Another strategy employed to test the reliability of the TEP response has been to synthetize the full waveform into a single index and then test the reliability of this index across multiple time points. For example, Casarotto et al. (2010) developed a nonparametric-permutation-based index to synthetize the degree of diversity between two TEPs at different time points in a single value, which they called the divergence index (DI), and used the receiver operating characteristics (ROC) of this value to capture the overall percentage of sensitivity (true positive rate, most similar to responsiveness in the Mokkink et al. 2010 framework, that a value changes when it should) and specificity (true negative rate, most similar to reliability in the Mokkink et al. 2010 framework, that a value remains the same when it should) of TEPs. They report a 96.7% sensitivity and 100% specificity, leading to two conclusions: first, that TEPs are not stereotypical and thus sensitive to stimulation parameter changes, including site, angle, or stimulation intensity; second, that TEPs are non-random, stable over time when parameters are held constant.

Another example is Ozdemir et al. (2020), who investigated the similarity of TEP responses at parietal, motor, and frontal sites, by computing the cosine similarity index between TEP matrices in different populations and at different time intervals. The similarity matrix containing the similarity index of all subjects two visits on separate days was then used to compute several similarity metrics used to capture reliability. They concluded that TEPs are highly reliable within individuals over time, but that they are highly heterogenous between individuals.

Other studies have tested the reliability of TEP-derived measures such as TMS–EEG long-interval cortical inhibition (Farzan et al. 2010), the symbolic transfer of entropy and vector autoregression (Ye et al. 2019), and interhemispheric signal propagation and balance (Casula et al. 2021). Except for vector autoregression, all the other indices appear reliable across repeated measures.

## 4. Criticisms of TMS–EEG-based measure reliability assessments to date

The literature on TMS–EEG-based measure reliability has consistently found an absence of significant differences between measures at different time points, as well as high correlation and high relative reliability of later TEP responses (>80 ms). However, only a minority of studies report relative or absolute reliability in a strict sense. Moreover, an accumulating amount of evidence indicates how those late TEP components that appear stable are, at least partially, contaminated with auditory and/or other sensory evoked potentials (ter Braack, de Vos, and van Putten 2015; Conde et al. 2019; Gordon et al. 2018; Nikouline, Ruohonen, and Ilmoniemi 1999; Rocchi et al. 2021; Tiitinen et al. 1999). These peripherally evoked responses are time-locked to the TMS pulse and are often highly consistent within-subjects, confounding reliability estimates. Therefore, caution should be taken when attempting to use those late latency responses as genuine, cortical-tissue-based, TEPs.

Overall, high relative reliability (ICC or CCC) is reported for TEPs. However, as discussed in section 2, this kind of reliability is relative to the population examined more than the technique used. Since the current literature on the relative reliability of TEPs has been built on studies involving young healthy adults, the generalizability of these findings is likely restricted to this population, with poor generalizability to other populations of interest, such as children, healthy older adults, or patient populations.

### *4.1. Methodological issues for TMS–EEG that impact reliability*

Early components of TEPs are by far the most variable, partly due to their high frequency and focality, but also because they are the most affected by recoding conditions and TMS-induced artifacts such as TMS-muscular and decay artifacts (Farzan and Bortoletto 2022; Hernandez-Pavon et al. 2022; Mutanen et al. 2016; Rogasch et al. 2017; Salo et al. 2020). The variability of these artifacts, as well as preprocessing strategies used among researchers to remove them, contribute to the difficulty in achieving high reliability.

In the raw signal, these artifacts can be an order of magnitude higher than the neuronal signal and require the use of sophisticated offline mathematical techniques to remove or attenuate them. Most of the available methodologies employ independent component analysis (ICA) to remove these artifacts (Atluri et al. 2016; Rogasch et al. 2017; Wu et al. 2018). However, ICA needs a subjective choice requiring expertise to determine what the signal is to spare and what the artifact is to remove, which is often not straightforward, introducing experimenter variability (Hernandez-Pavon et al. 2022). In addition, many TMS-related artifacts are time-locked to the TMS pulse, like the genuine cortical response. This may break the assumption of independency between the signal of interest and artifacts when using ICA to clean the EEG signal (Metsomaa, Sarvas, and Ilmoniemi 2014). Other algorithms that require fewer subjective choices have been recently published to remove those artifacts and are a promising alternative to the ICA approach (Mutanen et al. 2018).

Given the highly problematic nature of artefacts in TMS-EEG signal and the outlined limitations of preprocessing procedures, one strategy to reduce the impact of artefacts on TEP reliability may be to increase SNR in the recording phase. With this aim, Casarotto et al. (2022) suggested that visually checking the signal online and moving the coil based on a graphic-user interface allow to find stimulation parameters (e.g., site, coil orientation, intensity) that minimize artifacts and maximize signal. This approach may reduce the impact of preprocessing on the final signal (Casarotto et al. 2022), but it has not been tested yet if it may introduce additional variability when measuring TEP components.

### 4.2 Using t-tests and correlations to assess reliability

Using ANOVAs and *t*-tests to test for differences between measures taken from the same subjects in different sessions can be employed as the first step of reliability assessment (Bertazzoli et al. 2021; ter Braack et al. 2019; Corneal, Butler, and Wolf 2005; de Goede et al.

2020; Lioumis et al. 2009; Mancuso et al. 2021; Wolf et al. 2004). However, it needs to be considered that non-significant results imply neither good relative nor absolute reliability because the underlying hypothesis of these analyses refers to the mean of a distribution and does not refer to subject variance or residual error of a measure. As defined above, reliability (both absolute and relative) concerns the ability of a measure to remain stable in unchanging individuals. Instead, testing for differences using a *t*-test or an ANOVA is blind to the proportion in which a subject's measure changed with respect to the other subjects after repeated measures, if the group means of the repeated measures are not significantly different.

A demonstration of this case is provided in Table 2 (panel B) and visualized in Figure 1 (bottom right panel). In this situation, the responses change quite drastically within-subjects between time points, T0 and T1. In fact, both the *ICC* and *SEMeas*/*SDC* suggest low reliability. However, if we would just look at the result of the ANOVA, we will notice a *p* value of 0.08, and assuming a *p* threshold of 0.05 for statistical significance, we might conclude that data are not significantly different and might thus assume that the measurement is reliable. But this conclusion would be misguided.

Another common strategy used to assess reliability of TMS-EEG measures is through the calculation of Pearson's or Spearman's correlation coefficients (Balslev et al. 2007) after repeated measurements of the same index (Bertazzoli et al. 2021; Lioumis et al. 2009; Momi et al. 2021; Ozdemir et al. 2020; Ye et al. 2019). A common example could be the correlation of peaks' latency taken in two equal sessions to establish the reliability of that index. In contrast to the ICC, Pearson's and Spearman's correlations test whether a linear (or monotonic in the case of Spearman's) relationship exists between two variables, and are thus supposed to capture whether scores from test and retest sessions change in the same direction. In other words, whether all scores decrease or increase together. While Pearson's or Spearman's correlation coefficients can sometimes mimic the behavior of the ICC in some cases, they can lead to radically different results in other scenarios. For example, when most of the individual scores change consistently in one direction, Pearson's and Spearman's *statistics* can in high correlation when the reliability is actually be low (see Figure 2 and Table 3 for a demonstration).

## 5. Improving TMS–EEG-based measure reliability assessments

To fully appreciate the potential of TMS–EEG-based measures for clinical implementation, an extensive assessment of relative and absolute reliability of these measures is still needed. For this, more studies should collect repeated measurements of TMS–EEG data in different target populations to test relative (ICC) and absolute (SEMeas/SDC) reliability of the signal. Expanding our knowledge on TMS–EEG relative reliability in different populations, in which the variability of the tested population has the most effect, is fundamental for the diagnostic utility of TMS-EEG. At the same time, more data on absolute reliability in different TMS–EEG experimental setups need to be collected to establish the potential of TMS-EEG measures as prognostic biomarkers. More data on TMS-EEG absolute reliability would also allow to assess which TMS-EEG setup, recording procedure or apparatus and preprocessing method, is able to yield the highest absolute reliability, helping the field in moving toward the best practices.

Appropriate reliability tests like ICC and SEMeas or SDC should be implemented in future studies. Many studies used tests like Person's/Spearman's or cluster-based permutation *t*-tests to account for the complexity of the TMS-EEG signal as they allow for an unbiased test of the whole TEP response without losing signal dimensions. Indeed, ICC and SEMeas or SDC have been usually calculated when a single numeric value is extracted in each individual, e.g.*,* peak amplitude, possibly due to high computational demands. However, reducing the TEPs to their peak amplitude and latency may add a further layer of variability since the process of peak extraction itself is not straightforward and the methodologies are not standardized (Luck 2014).

To tackle this issue, future studies should report the ICC and SEMeas/SDC computed at each time point and electrode. This strategy would allow to follow the changes in reliability across time and electrodes, allowing a continuous evaluation of the most reliable intervals of the TMS–EEG response in both time and space.

Here we support the feasibility of this approach using a dataset with stimulation on left IPL in young healthy subjects (for details on the data please refer to the original paper Esposito et al. (2022). The dataset is publicly available at https://gin.g-node.org/CIMeC/TMS-EEG_brain_connectivity_BIDS/src/master in BIDS standard) (Gorgolewski et al. 2016; Pernet et al. 2019). Figure 3 and Figure 4 show the ICC and the SDC, respectively, over several channels. It is possible to notice that the ICC remains 'moderate' to 'substantial' based on the Shrout 1998 standard at >0.8 in most of the electrodes after 150 ms from the TMS pulse, except for electrode F3, in favor of the previous finding of high reliability of late TEP latencies. Accordingly, early latencies show instead mixed results in terms of ICC, with contralateral electrodes (P4, C4, F4), vertex (Cz) and electrode nearest to the stimulation (P3) stabilizing earlier than the others (Figure 3). The continuous SDC, instead, shows a general high absolute reliability at approximately 50 ms and 200 ms (SDC ~3 µV), except for electrode Cz for the late latency (Figure 4). Therefore, without an a priori hypothesis for peak extraction, we would already know that latencies of approximately 50 ms and 200 ms would represent windows of high relative and absolute reliability at specific electrodes.

### *5.1 From bench to bedside: conventional evoked potentials in clinical diagnostics*

Evoked potentials (EPs) and event-related potentials (ERPs) are often used in clinical neurophysiology and neurology diagnostics and prognostics (Duncan et al. 2009; Kappenman and Luck 2011). They also have clear relevance in intraoperative monitoring (MacDonald et al.

2013; McKhann et al. 2011; Nuwer et al. 2012; Oh et al. 2012; Wong et al. 2022). The pathways of the EPs and ERPs from bench to bedside and standard clinical practice have undergone a rigorous validation with animal, stimulation, pilot, clinical, and meta-analysis, studies. With these EPs, their clinical application has advanced due to gained experience, interest, and innovation.

In clinical practice, the common EPs that are used in diagnostics are, for example, MEPs, somatosensory evoked potentials (SEPs), auditory evoked potentials (AEPs) and visual evoked potentials (VEPs) (Celesia et al. 1993; Chen et al. 2008b; Cruccu et al. 2008; Holder et al. 2010; Karl E. Misulis 2001; Odom et al. 2016). These diagnostics are clinically useful because they have been well mapped out in different populations. They thus are known to remain stable in unchanging conditions (reliability) and allow for reliable detection of neurophysiological changes (responsiveness). We have clear specifications about normal or healthy ranges for EP latencies and amplitudes, as well as well-established ranges for normal and intraindividual and interhemispheric differences in latencies and amplitudes (Husain 2021; Nuwer et al. 1994; Sonoo et al. 1996). Through normative data and the associated statistics, we can thus identify the abnormal, motivating their clinical application.

In parallel with the definition of normative data, which is usually done with large-sample cross-sectional studies separately for each target population, reliability assessment of EPs played a crucial role in determining their clinical utility (see recent examples Clayson et al. 2021; Cunningham et al. 2023; Giuffre et al. 2021; Jetha, Segalowitz, and Gatzke-Kopp 2021; Nazarova et al. 2021; Therrien-Blanchet et al. 2022; Vernillo et al. 2022; for review see Beaulieu et al. 2017; Schambra et al. 2015).

While establishing normative ranges of between-subject variability might help a clinician draw lines between physiology and pathology, reliability assessments are what determines whether a measure is fit for clinical use as a diagnostic or prognostic tool or not. Without well-documented and standardized measures of reliability in studies using these measures from basic scientific studies, basic scientific research cannot inform or benefit clinicians. This is the core motivation of standardizing reliability estimates more generally, and here, with TMS-EEG. TMS-EEG-based measures like TEPs have potential to be diagnostic and prognostic clinical tools. However, just has been done before with EPs and ERPs, extensive reliability evidence is needed first. An agreement on the analyzed and quantified responses should be reached, and establishing a reliability assessment or set of reliability assessments that enables the establishment of normative values are the next steps.

*5.2 Contributions of multicentric experiments and data sharing to reliability assessments*

TMS–EEG experiments are long and expensive, especially when an MRI is collected to ensure precise and accurate stimulation with neuronavigational systems (Julkunen et al. 2009). Moreover, testing fragile populations such as older adults or patient populations, can be difficult and sometimes infeasible, considering that each subject might need to be tested multiple times. These issues often hinder the collection of sufficiently large datasets that are needed for a thorough assessment of reliability. One solution that was successfully applied in other neuroimaging fields to improve replicability more broadly is the contemporaneous collection of the same data across many labs or institutions (Pavlov et al. 2021; Weiner et al. 2017). This approach increases the generalizability of the findings and allows the collection of large datasets that can be made available for the field at large. Big data collection not only helps in assessing reliability of TMS-EEG-based measures, but also allows establishing normative values for physiological and pathological TMS-EEG-based measures needed for translation into clinical practice. An initiative with this aim has recently been launched for TMS-EEG ("Team for TMS-EEG" T4TE, Bortoletto et al. 2022).

TMS–EEG signals are affected by preprocessing choices (Bertazzoli et al. 2021), which is one potential source of variability between labs and poses a challenge for reliability. Testing the reliability of the same TMS–EEG signal after different preprocessing (intrarater reliability (Beaulieu et al. 2017; Mokkink et al. 2010)) will also be fundamental for assessing the potential of the technique for diagnostic and prognostic uses. Here, the most efficient way to evaluate reliability is by allowing open access to the raw data to assess the impact of the preprocessing choices on several conditions of data acquisition. To be effective, data sharing shall follow the F.A.I.R. principles (www.go-fair.org), as it was successfully implemented in the brain-imaging data structure (BIDS) initiative (Gorgolewski et al. 2016; Pernet et al. 2019) for other neuroimaging modalities. Moreover, the BIDS initiative is only just beginning to provide a clear specification on how to describe concurrent TMS and EEG, among other non-invasive stimulation techniques (see BIDS-extension proposal for non-invasive brain stimulation experiments bids.neuroimaging.io/get_involved.html).

## 6. Conclusions

This review focuses on the assessment of relative and absolute reliability for TMS-EEG-based measures, relevant for the clinical field as diagnostic and prognostic biomarkers. The current literature highlights the absence of significant differences between TEPs in a test-retest fashion. Additionally, test-retest TEPs show an increasing trend of correlation from early to late latency. Some studies have found high relative reliability and low absolute reliability for late latency TEPs. In general, the reliability assessment of TMS-EEG measures is mainly restricted to the young, healthy population. Importantly, the assessment of absolute reliability has been often neglected, despite being an important property for establishing the potential of a TMS-EEG related measure as a prognostic biomarker. Therefore, for TMS-EEG to enter clinical practice, an extensive assessment of relative and absolute reliability of TMS-EEG related measures in different populations and experimental settings, is needed. Ongoing multicentric studies and standardized data sharing will be key to establish and improving reliability of TMS-EEG signal.

**Acknowledgements**

Marta Bortoletto acknowledges fundings from the Italian Ministry of Health ("Bando della ricerca finalizzata 2016 - Giovani Ricercatori" grant no: GR-2016-02364132 and Ricerca Corrente). Petro Julkunen acknowledges funding from the Academy of Finland (grant no: 322423).

**Authors Contribution**

Giacomo Bertazzoli: *Conceptualization, Investigation, Writing - Original Draft, Writing - Review & Editing, Visualization.*

Carlo Miniussi: *Writing - Review & Editing, Supervision.*

Petro Julkunen: *Writing - Review & Editing.*

Marta Bortoletto: *Conceptualization, Writing - Review & Editing, Supervision, Project administration, Funding acquisition.*

**Declaration of competing interests**

Petro Julkunen shares an unrelated patent with, and has received consulting fees from Nexstim Plc, Helsinki, Finland, manufacturer of navigated TMS systems. Other authors declare no competing interests.

# References

Arai, Naohiro, Tomoya Nakanishi, Shinichiro Nakajima, Xuemei Li, Masataka Wada, Zafiris J. Daskalakis, Michelle S. Goodman, Daniel M. Blumberger, Masaru Mimura, and Yoshihiro Noda. 2021. "Insights of Neurophysiology on Unconscious State Using Combined Transcranial Magnetic Stimulation and Electroencephalography: A Systematic Review." *Neuroscience and Biobehavioral Reviews* 131:293–312.

Atkinson, Greg, and Alan M. Nevill. 1998. "Statistical Methods for Assessing Measurement Error (Reliability) in Variables Relevant to Sports Medicine." *Sports Medicine* 26(4):217–38.

Atluri, Sravya, Matthew Frehlich, Ye Mei, Luis Garcia Dominguez, Nigel C. Rogasch, Willy Wong, Zafiris J. Daskalakis, and Faranak Farzan. 2016. "TMSEEG: A MATLAB-Based Graphical User Interface for Processing Electrophysiological Signals during Transcranial Magnetic Stimulation." *Frontiers in Neural Circuits* 10:78.

Bagattini, Chiara, Tuomas P. Mutanen, Claudia Fracassi, Rosa Manenti, Maria Cotelli, Risto J. Ilmoniemi, Carlo Miniussi, and Marta Bortoletto. 2019. "Predicting Alzheimer's Disease Severity by Means of TMS–EEG Coregistration." *Neurobiology of Aging* 80:38–45.

Bai, Yang, Xiaoyu Xia, Jiannan Kang, Xiaoxiao Yin, Yi Yang, Jianghong He, and Xiaoli Li. 2016. "Evaluating the Effect of Repetitive Transcranial Magnetic Stimulation on Disorders of Consciousness by Using TMS-EEG." *Frontiers in Neuroscience* 10(OCT).

Balslev, Daniela, Wouter Braet, Craig McAllister, and R. Chris Miall. 2007. "Between-subject Variability in Optimal Current Direction for Transcranial Magnetic Stimulation of the Motor Cortex." *Journal of Neuroscience Methods* 162(1–2):309–13.

Beaulieu, Louis David, Véronique H. Flamand, Hugo Massé-Alarie, and Cyril Schneider. 2017. "Reliability and Minimal Detectable Change of Transcranial Magnetic Stimulation Outcomes in Healthy Adults: A Systematic Review." *Brain Stimulation* 10(2):196–213.

Bertazzoli, Giacomo, Romina Esposito, Tuomas P. Mutanen, Clarissa Ferrari, Risto J. Ilmoniemi, Carlo Miniussi, and Marta Bortoletto. 2021. "The Impact of Artifact Removal Approaches on TMS-EEG Signal." *NeuroImage* 239:118272.

Bodart, Olivier, Olivia Gosseries, Sarah Wannez, Aurore Thibaut, Jitka Annen, Melanie Boly, Mario Rosanova, Adenauer G. Casali, Silvia Casarotto, Giulio Tononi, Marcello Massimini, and Steven Laureys. 2017. "Measures of Metabolism and Complexity in the Brain of Patients with Disorders of Consciousness." *NeuroImage: Clinical* 14:354–62.

Bonato, C., C. Miniussi, and P. M. Rossini. 2006. "Transcranial Magnetic Stimulation and Cortical Evoked Potentials: A TMS/EEG Co-Registration Study." *Clinical Neurophysiology* 117(8):1699–1707.

Borich, Michael R., Lewis A. Wheaton, Sonia M. Brodie, Bimal Lakhani, and Lara A. Boyd. 2016. "Evaluating Interhemispheric Cortical Responses to Transcranial Magnetic Stimulation in Chronic Stroke: A TMS-EEG Investigation." *Neuroscience Letters* 618:25–30.

Bortoletto, Marta, Domenica Veniero, Petro Julkunen, Julio C. Hernandez-Pavon, Tuomas P. Mutanen, Agnese Zazio, and Chiara Bagattini. 2022. "T4TE: Team for TMS−EEG and Improve Reproducibility through an Open Collaborative Initiative." *Brain Stimulation: Basic, Translational, and Clinical Research in Neuromodulation* 0(0).

Bortoletto, Marta, Domenica Veniero, Gregor Thut, and Carlo Miniussi. 2015. "The Contribution of TMS-EEG Coregistration in the Exploration of the Human Cortical Connectome." *Neuroscience and Biobehavioral Reviews* 49:114–24.

ter Braack, Esther M., Annika A. de Goede, and Michel J. A. M. van Putten. 2019. "Resting Motor Threshold, MEP and TEP Variability During Daytime." *Brain Topography* 32(1):17–27.

ter Braack, Esther M., Cecile C. de Vos, and Michel J. A. M. van Putten. 2015. "Masking the Auditory Evoked Potential in TMS–EEG: A Comparison of Various Methods." *Brain Topography* 28(3):520–28.

Canali, Paola, Simone Sarasso, Mario Rosanova, Silvia Casarotto, Giovanna Sferrazza-Papa, Olivia Gosseries, Matteo Fecchio, Marcello Massimini, Maurizio Mariotti, Roberto Cavallaro, Enrico Smeraldi, Cristina Colombo, and Francesco Benedetti. 2015. "Shared Reduction of Oscillatory Natural Frequencies in Bipolar Disorder, Major Depressive Disorder and Schizophrenia." *Journal of Affective Disorders* 184:111–15.

Carrasco, Josep L., and Lluís Jover. 2003. "Estimating the Generalized Concordance Correlation Coefficient through Variance Components." *Biometrics* 59(4):849–58.

Casarotto, Silvia, Angela Comanducci, Mario Rosanova, Simone Sarasso, Matteo Fecchio, Martino Napolitani, Andrea Pigorini, Adenauer G. Casali, Pietro D. Trimarchi, Melanie Boly, Olivia Gosseries, Olivier Bodart, Francesco Curto, Cristina Landi, Maurizio Mariotti, Guya Devalle, Steven Laureys, Giulio Tononi, and Marcello Massimini. 2016. "Stratification of Unresponsive Patients by an Independently Validated Index of Brain Complexity." *Annals of Neurology* 80(5):718–29.

Casarotto, Silvia, Matteo Fecchio, Mario Rosanova, Giuseppe Varone, Sasha D'Ambrosio, Simone Sarasso, Andrea Pigorini, Simone Russo, Angela Comanducci, Risto J. Ilmoniemi, and Marcello Massimini. 2022. "The Rt-TEP Tool: Real-Time Visualization of TMS-Evoked Potentials to Maximize Cortical Activation and Minimize Artifacts." *Journal of Neuroscience Methods* 370.

Casarotto, Silvia, Leonor J. Romer. Lauro, Valentina Bellina, Adenauer G. Casali, Mario Rosanova, Andrea Pigorini, Stefano Defendi, Maurizio Mariotti, and Marcello Massimini. 2010. "EEG Responses to TMS Are Sensitive to Changes in the Perturbation Parameters and Repeatable over Time." *PLoS ONE* 5(4).

Casarotto, Silvia, Sara Määttä, Sanna Kaisa Herukka, Andrea Pigorini, Martino Napolitani, Olivia Gosseries, Eini Niskanen, Mervi Könönen, Esa Mervaala, Mario Rosanova, Hilkka Soininen, and Marcello Massimini. 2011. "Transcranial Magnetic Stimulation-Evoked EEG/Cortical Potentials in Physiological and Pathological Aging." *NeuroReport* 22(12):592–97.

Casula, Elias Paolo, Gaetano Tieri, Lorenzo Rocchi, Rachele Pezzetta, Michele Maiella, Enea Francesco Pavone, Salvatore Maria Aglioti, and Giacomo Koch. 2022. "Feeling of Ownership over an Embodied Avatar's Hand Brings About Fast Changes of Fronto-Parietal Cortical Dynamics." *The Journal of Neuroscience : The Official Journal of the Society for Neuroscience* 42(4):692–701.

Chen, Robert, Didier Cros, Antonio Curra, Vincenzo Di Lazzaro, Jean Pascal Lefaucheur, Michel R. Magistris, Kerry Mills, Kai M. Rösler, William J. Triggs, Yoshikazu Ugawa, and Ulf Ziemann. 2008. "The Clinical Diagnostic Utility of Transcranial Magnetic Stimulation: Report of an IFCN Committee." *Clinical Neurophysiology* 119(3):504–32.

Cipollari, Susanna, Domenica Veniero, Carmela Razzano, Carlo Caltagirone, Giacomo Koch, and Paola Marangolo. 2015. "Combining TMS-EEG with Transcranial Direct Current Stimulation Language Treatment in Aphasia." *Expert Review of Neurotherapeutics* 15(7):833–45.

Conde, Virginia, Leo Tomasevic, Irina Akopian, Konrad Stanek, Guilherme B. Saturnino, Axel

Thielscher, Til Ole Bergmann, and Hartwig Roman Siebner. 2019. "The Non-Transcranial TMS-Evoked Potential Is an Inherent Source of Ambiguity in TMS-EEG Studies." *NeuroImage* 185:300–312.

Corneal, Scott F., Andrew J. Butler, and Steven L. Wolf. 2005. "Intra- and Intersubject Reliability of Abductor Pollicis Brevis Muscle Motor Map Characteristics with Transcranial Magnetic Stimulation." *Archives of Physical Medicine and Rehabilitation* 86(8):1670–75.

D'Agati, Elisa, Thomas Hoegl, Gabriel Dippel, Paolo Curatolo, Stephan Bender, Oliver Kratz, Gunther H. Moll, and Hartmut Heinrich. 2014. "Motor Cortical Inhibition in ADHD: Modulation of the Transcranial Magnetic Stimulation-Evoked N100 in a Response Control Task." *Journal of Neural Transmission (Vienna, Austria : 1996)* 121(3):315–25.

Esposito, Romina, Marta Bortoletto, Domenico Zacà, Paolo Avesani, and Carlo Miniussi. 2022. "An Integrated TMS-EEG and MRI Approach to Explore the Interregional Connectivity of the Default Mode Network." *Brain Structure & Function* 227(3):1133–44.

Farzan, Faranak, Mera S. Barr, Andrea J. Levinson, Robert Chen, Willy Wong, Paul B. Fitzgerald, and Zafiris J. Daskalakis. 2010. "Reliability of Long-Interval Cortical Inhibition in Healthy Human Subjects: A TMS-EEG Study." *Journal of Neurophysiology* 104(3):1339–46.

Farzan, Faranak, and Marta Bortoletto. 2022. "Identification and Verification of a 'true' TMS Evoked Potential in TMS-EEG." *Journal of Neuroscience Methods* 378:109651.

Ferrarelli, Fabio, Marcello Massimini, Simone Sarasso, Adenauer Casali, Brady A. Riedner, Giuditta Angelini, Giulio Tononi, and Robert A. Pearce. 2010. "Breakdown in Cortical Effective Connectivity during Midazolam-Induced Loss of Consciousness." *Proceedings of the National Academy of Sciences of the United States of America* 107(6):2681–86.

Ferreri, Florinda, Andrea Guerra, Luca Vollero, David Ponzo, Sara Määtta, Mervi Könönen, Fabrizio Vecchio, Patrizio Pasqualetti, Francesca Miraglia, Ilaria Simonelli, Maurizio Corbetta, and Paolo Maria Rossini. 2021. "TMS-EEG Biomarkers of Amnestic Mild Cognitive Impairment Due to Alzheimer's Disease: A Proof-of-Concept Six Years Prospective Study." *Frontiers in Aging Neuroscience* 13.

Ferreri, Florinda, Fabrizio Vecchio, Luca Vollero, Andrea Guerra, Sara Petrichella, David Ponzo, Sara Määtta, Esa Mervaala, Mervi Könönen, Francesca Ursini, Patrizio Pasqualetti, Giulio Iannello, Paolo Maria Rossini, and Vincenzo Di Lazzaro. 2016. "Sensorimotor Cortex Excitability and Connectivity in Alzheimer's Disease: A TMS-EEG Co-Registration Study." *Human Brain Mapping* 37(6):2083–96.

Formaggio, Emanuela, Marianna Cavinato, Silvia Francesca Storti, Paolo Tonin, Francesco Piccione, and Paolo Manganotti. 2016. "Assessment of Event-Related EEG Power After Single-Pulse TMS in Unresponsive Wakefulness Syndrome and Minimally Conscious State Patients." *Brain Topography* 29(2):322–33.

Giraudeau, B. 1996. "Negative Values of the Intraclass Correlation Coefficient Are Not Theoretically Possible." *Journal of Clinical Epidemiology* 49(10):1205–6.

de Goede, Annika A., Irene Cumplido-Mayoral, and Michel J. A. M. van Putten. 2020. "Spatiotemporal Dynamics of Single and Paired Pulse TMS-EEG Responses." *Brain Topography* 33(4):425–37.

Gordon, Pedro Caldana, Debora Desideri, Paolo Belardinelli, Christoph Zrenner, and Ulf Ziemann. 2018. "Comparison of Cortical EEG Responses to Realistic Sham versus Real TMS of Human Motor Cortex." *Brain Stimulation* 11(6):1322–30.

Gorgolewski, Krzysztof J., Tibor Auer, Vince D. Calhoun, R. Cameron Craddock, Samir Das, Eugene P. Duff, Guillaume Flandin, Satrajit S. Ghosh, Tristan Glatard, Yaroslav O.

Halchenko, Daniel A. Handwerker, Michael Hanke, David Keator, Xiangrui Li, Zachary Michael, Camille Maumet, B. Nolan Nichols, Thomas E. Nichols, John Pellman, Jean Baptiste Poline, Ariel Rokem, Gunnar Schaefer, Vanessa Sochat, William Triplett, Jessica A. Turner, Gaël Varoquaux, and Russell A. Poldrack. 2016. "The Brain Imaging Data Structure, a Format for Organizing and Describing Outputs of Neuroimaging Experiments." *Scientific Data 2016 3:1* 3(1):1–9.

Gosseries, O., A. Thibaut, M. Boly, M. Rosanova, M. Massimini, and S. Laureys. 2014. "Assessing Consciousness in Coma and Related States Using Transcranial Magnetic Stimulation Combined with Electroencephalography." *Annales Francaises d'anesthesie et de Reanimation* 33(2):65–71.

Gosseries, Olivia, Simone Sarasso, Silvia Casarotto, Mélanie Boly, Caroline Schnakers, Martino Napolitani, Marie Aurélie Bruno, Didier Ledoux, Jean Flory Tshibanda, Marcello Massimini, Steven Laureys, and Mario Rosanova. 2015. "On the Cerebral Origin of EEG Responses to TMS: Insights From Severe Cortical Lesions." *Brain Stimulation* 8(1):142–49.

Hernandez-Pavon, Julio C., Dimitris Kugiumtzis, Christoph Zrenner, Vasilios K. Kimiskidis, and Johanna Metsomaa. 2022. "Removing Artifacts from TMS-Evoked EEG: A Methods Review and a Unifying Theoretical Framework." *Journal of Neuroscience Methods* 376.

Hernandez-Pavon Julio C., Domenica Veniero, Til Ole Bergmann, Paolo Belardinelli, Marta Bortoletto, Silvia Casarotto, Elias Casula, Faranak Farzan, Matteo Fecchio , Petro Julkunen, Elisa Kallioniemi, Pantelis Lioumis, Johanna Metsomaa, Carlo Miniussi, Tuomas P. Mutanen, Lorenzo Rocchi, Nigel C. Rogasch, Mouhsin M. Shafi, Hartwig R. Siebner, Gregor Thut, Christoph Zrenner, Ulf Ziemann, Risto J. Ilmoniemi. 2023. TMS Combined with EEG: Recommendations and Open Issues. *Brain Stimulation.* 16(2):567-593. doi: 10.1016/j.brs.2023.02.009.

Hopkins, Will G. 2000. "Measures of Reliability in Sports Medicine and Science." *Sports Medicine (Auckland, N.Z.)* 30(1):1–15.

Ilmoniemi, Risto J., and Dubravko Kičić. 2010. "Methodology for Combined TMS and EEG." *Brain Topography* 22(4):233–48.

Julkunen, Petro, Anne M. Jauhiainen, Mervi Knnen, Ari Pääkknen, Jari Karhu, and Hilkka Soininen. 2011. "Combining Transcranial Magnetic Stimulation and Electroencephalography May Contribute to Assess the Severity of Alzheimer's Disease." *International Journal of Alzheimer's Disease* 2011.

Julkunen, Petro, Anne M. Jauhiainen, Susanna Westerén-Punnonen, Eriikka Pirinen, Hilkka Soininen, Mervi Könönen, Ari Pääkkönen, Sara Määttä, and Jari Karhu. 2008. "Navigated TMS Combined with EEG in Mild Cognitive Impairment and Alzheimer's Disease: A Pilot Study." *Journal of Neuroscience Methods* 172(2):270–76.

Julkunen, Petro, Vasilios K. Kimiskidis, and Paolo Belardinelli. 2022. "Bridging the Gap: TMS-EEG from Lab to Clinic." *Journal of Neuroscience Methods* 369:109482.

Kähkönen, Seppo, Juha Wilenius, Soile Komssi, and Risto J. Ilmoniemi. 2004. "Distinct Differences in Cortical Reactivity of Motor and Prefrontal Cortices to Magnetic Stimulation." *Clinical Neurophysiology* 115(3):583–88.

Kallioniemi, Elisa, Jukka Saari, Florinda Ferreri, and Sara Määttä. 2022. "TMS-EEG Responses across the Lifespan: Measurement, Methods for Characterisation and Identified Responses." *Journal of Neuroscience Methods* 366:109430.

Kerwin, Lewis J., Corey J. Keller, Wei Wu, Manjari Narayan, and Amit Etkin. 2018. "Test-Retest Reliability of Transcranial Magnetic Stimulation EEG Evoked Potentials." *Brain Stimulation*

11(3):536–44.

King, Tonya S., Vernon M. Chinchilli, Kai Ling Wang, and Josep L. Carrasco. 2007. "A Class of Repeated Measures Concordance Correlation Coefficients." *Journal of Biopharmaceutical Statistics* 17(4):653–72.

Kirkovski, Melissa, Nigel C. Rogasch, Takashi Saeki, Bernadette M. Fitzgibbon, Peter G. Enticott, and Paul B. Fitzgerald. 2016. "Single Pulse Transcranial Magnetic Stimulation-Electroencephalogram Reveals No Electrophysiological Abnormality in Adults with High-Functioning Autism Spectrum Disorder." *Journal of Child and Adolescent Psychopharmacology* 26(7):606–16.

Kobayashi, Masahito, and Alvaro Pascual-Leone. 2003. "Transcranial Magnetic Stimulation in Neurology." *Lancet Neurology* 2(3):145–56.

Koch, Giacomo, Sonia Bonnì, Maria Concetta Pellicciari, Elias P. Casula, Matteo Mancini, Romina Esposito, Viviana Ponzo, Silvia Picazio, Francesco Di Lorenzo, Laura Serra, Caterina Motta, Michele Maiella, Camillo Marra, Mara Cercignani, Alessandro Martorana, Carlo Caltagirone, and Marco Bozzali. 2018. "Transcranial Magnetic Stimulation of the Precuneus Enhances Memory and Neural Activity in Prodromal Alzheimer's Disease." *NeuroImage* 169:302–11.

Kumar, Sanjeev, Reza Zomorrodi, Zaid Ghazala, Michelle S. Goodman, Daniel M. Blumberger, Amay Cheam, Corinne Fischer, Zafiris J. Daskalakis, Benoit H. Mulsant, Bruce G. Pollock, and Tarek K. Rajji. 2017. "Extent of Dorsolateral Prefrontal Cortex Plasticity and Its Association With Working Memory in Patients With Alzheimer Disease." *JAMA Psychiatry* 74(12):1266–74.

Levit-Binnun, Nava, Vladimir Litvak, Hillel Pratt, Elisha Moses, Menashe Zaroor, and Avi Peled. 2009. "Differences in TMS-Evoked Responses between Schizophrenia Patients and Healthy Controls Can Be Observed without a Dedicated EEG System." *Clinical Neurophysiology* 121:332–39.

Lin, Lawrence I. Kuei. 1989. "A Concordance Correlation Coefficient to Evaluate Reproducibility." *Biometrics* 45(1):255.

Lioumis, Pantelis, Dubravko Kičić, Petri Savolainen, Jyrki P. Mäkelä, and Seppo Kähkönen. 2009. "Reproducibility of TMS - Evoked EEG Responses." *Human Brain Mapping* 30(4):1387–96.

Luck, Steven J. (Steven John). 2014. *An Introduction to the Event-Related Potential Technique*. MIT Press.

Mancuso, Marco, Valerio Sveva, Alessandro Cruciani, Katlyn Brown, Jaime Ibáñez, Vishal Rawji, Elias Casula, Isabella Premoli, Sasha D'Ambrosio, John Rothwell, and Lorenzo Rocchi. 2021. "Transcranial Evoked Potentials Can Be Reliably Recorded with Active Electrodes." *Brain Sciences* 11(2):1–16.

Manganotti, Paolo, Michele Acler, Stefano Masiero, and Alessandra Del Felice. 2015. "TMS-Evoked N100 Responses as a Prognostic Factor in Acute Stroke." *Functional Neurology* 30(2):125–30.

Massimini, M., F. Ferrarelli, S. Sarasso, and G. Tononi. 2012. "Cortical Mechanisms of Loss of Consciousness: Insight from TMS/EEG Studies." *Archives Italiennes de Biologie* 150(2–3):44–55.

McManus, I. C. 2012. "The Misinterpretation of the Standard Error of Measurement in Medical Education: A Primer on the Problems, Pitfalls and Peculiarities of the Three Different Standard Errors of Measurement." Pp. 569–76 in *Medical Teacher*. Vol. 34. Med Teach.

Metsomaa, Johanna, Jukka Sarvas, and Risto J. Ilmoniemi. 2014. "Multi-Trial Evoked EEG and Independent Component Analysis." *Journal of Neuroscience Methods* 228:15–26.

Mokkink, Lidwine B., Caroline B. Terwee, Donald L. Patrick, Jordi Alonso, Paul W. Stratford, Dirk L. Knol, Lex M. Bouter, and Henrica C. W. de Vet. 2010. "The COSMIN Study Reached International Consensus on Taxonomy, Terminology, and Definitions of Measurement Properties for Health-Related Patient-Reported Outcomes." *Journal of Clinical Epidemiology* 63(7):737–45.

Momi, Davide, Recep A. Ozdemir, Ehsan Tadayon, Pierre Boucher, Alberto Di Domenico, Mirco Fasolo, Mouhsin M. Shafi, Alvaro Pascual-Leone, and Emiliano Santarnecchi. 2021. "Perturbation of Resting-State Network Nodes Preferentially Propagates to Structurally Rather than Functionally Connected Regions." *Scientific Reports* 11(1).

Momi, Davide, Recep A. Ozdemir, Ehsan Tadayon, Pierre Boucher, Mouhsin M. Shafi, Alvaro Pascual-Leone, and Emiliano Santarnecchi. 2021. "Network-Level Macroscale Structural Connectivity Predicts Propagation of Transcranial Magnetic Stimulation." *NeuroImage* 229.

Mutanen, Tuomas P., Matleena Kukkonen, Jaakko O. Nieminen, Matti Stenroos, Jukka Sarvas, and Risto J. Ilmoniemi. 2016. "Recovering TMS-Evoked EEG Responses Masked by Muscle Artifacts." *NeuroImage* 139:157–66.

Mutanen, Tuomas P., Johanna Metsomaa, Sara Liljander, and Risto J. Ilmoniemi. 2018. "Automatic and Robust Noise Suppression in EEG and MEG: The SOUND Algorithm." *NeuroImage* 166:135–51.

Naim-Feil, Jodie, John L. Bradshaw, Nigel C. Rogasch, Zafiris J. Daskalakis, Dianne M. Sheppard, Dan I. Lubman, and Paul B. Fitzgerald. 2016. "Cortical Inhibition within Motor and Frontal Regions in Alcohol Dependence Post-Detoxification: A Pilot TMS-EEG Study." *The World Journal of Biological Psychiatry : The Official Journal of the World Federation of Societies of Biological Psychiatry* 17(7):547–56.

Nikouline, V., Jarmo Ruohonen, and Risto J. Ilmoniemi. 1999. "The Role of the Coil Click in TMS Assessed with Simultaneous EEG." *Clinical Neurophysiology* 110(8):1325–28.

Noda, Yoshihiro, Mera S. Barr, Reza Zomorrodi, Robin F. H. Cash, Tarek K. Rajji, Faranak Farzan, Robert Chen, Tony P. George, Zafiris J. Daskalakis, and Daniel M. Blumberger. 2018. "Reduced Short-Latency Afferent Inhibition in Prefrontal but Not Motor Cortex and Its Association With Executive Function in Schizophrenia: A Combined TMS-EEG Study." *Schizophrenia Bulletin* 44(1):193–202.

Noda, Yoshihiro, Reza Zomorrodi, Fidel Vila-Rodriguez, Jonathan Downar, Faranak Farzan, Robin F. H. Cash, Tarek K. Rajji, Zafiris J. Daskalakis, and Daniel M. Blumberger. 2018. "Impaired Neuroplasticity in the Prefrontal Cortex in Depression Indexed through Paired Associative Stimulation." *Depression and Anxiety* 35(5):448–56.

Oostenveld, Robert, Pascal Fries, Eric Maris, and Jan Mathijs Schoffelen. 2011. "FieldTrip: Open Source Software for Advanced Analysis of MEG, EEG, and Invasive Electrophysiological Data." *Computational Intelligence and Neuroscience* 2011:1–9.

Ozdemir, Recep A., Ehsan Tadayon, Pierre Boucher, Davide Momi, Kelly A. Karakhanyan, Michael D. Fox, Mark A. Halko, Alvaro Pascual-Leone, Mouhsin M. Shafi, and Emiliano Santarnecchi. 2020. "Individualized Perturbation of the Human Connectome Reveals Reproducible Biomarkers of Network Dynamics Relevant to Cognition." *Proceedings of the National Academy of Sciences of the United States of America* 117(14):8115–25.

Pavlov, Yuri G., Nika Adamian, Stefan Appelhoff, Mahnaz Arvaneh, Christopher S. Y. Benwell, Christian Beste, Amy R. Bland, Daniel E. Bradford, Florian Bublatzky, Niko A. Busch, Peter E. Clayson, Damian Cruse, Artur Czeszumski, Anna Dreber, Guillaume Dumas, Benedikt Ehinger, Giorgio Ganis, Xun He, José A. Hinojosa, Christoph Huber-Huber, Michael

Inzlicht, Bradley N. Jack, Magnus Johannesson, Rhiannon Jones, Evgenii Kalenkovich, Laura Kaltwasser, Hamid Karimi-Rouzbahani, Andreas Keil, Peter König, Layla Kouara, Louisa Kulke, Cecile D. Ladouceur, Nicolas Langer, Heinrich R. Liesefeld, David Luque, Annmarie MacNamara, Liad Mudrik, Muthuraman Muthuraman, Lauren B. Neal, Gustav Nilsonne, Guiomar Niso, Sebastian Ocklenburg, Robert Oostenveld, Cyril R. Pernet, Gilles Pourtois, Manuela Ruzzoli, Sarah M. Sass, Alexandre Schaefer, Magdalena Senderecka, Joel S. Snyder, Christian K. Tamnes, Emmanuelle Tognoli, Marieke K. van Vugt, Edelyn Verona, Robin Vloeberghs, Dominik Welke, Jan R. Wessel, Ilya Zakharov, and Faisal Mushtaq. 2021. "#EEGManyLabs: Investigating the Replicability of Influential EEG Experiments." *Cortex; a Journal Devoted to the Study of the Nervous System and Behavior* 144:213–29.

Pellicciari, Maria Concetta, Sonia Bonnì, Viviana Ponzo, Alex Martino Cinnera, Matteo Mancini, Elias Paolo Casula, Fabrizio Sallustio, Stefano Paolucci, Carlo Caltagirone, and Giacomo Koch. 2018. "Dynamic Reorganization of TMS-Evoked Activity in Subcortical Stroke Patients." *NeuroImage* 175:365–78.

Pernet, Cyril R., Stefan Appelhoff, Krzysztof J. Gorgolewski, Guillaume Flandin, Christophe Phillips, Arnaud Delorme, and Robert Oostenveld. 2019. "EEG-BIDS, an Extension to the Brain Imaging Data Structure for Electroencephalography." *Scientific Data* 6(1):1–5.

Ragazzoni, Aldo, Massimo Cincotta, Fabio Giovannelli, Damian Cruse, Bryan Young, Carlo Miniussi, and Simone Rossi. 2017. "Clinical Neurophysiology of Prolonged Disorders of Consciousness: From Diagnostic Stimulation to Therapeutic Neuromodulation." *Clinical Neurophysiology* 128:1629–46.

Ragazzoni, Aldo, Cornelia Pirulli, Domenica Veniero, Matteo Feurra, Massimo Cincotta, Fabio Giovannelli, Roberta Chiaramonti, Mario Lino, Simone Rossi, and Carlo Miniussi. 2013. "Vegetative versus Minimally Conscious States: A Study Using TMS-EEG, Sensory and Event-Related Potentials." *PloS One* 8(2).

Rocchi, Lorenzo, Alessandro Di Santo, Katlyn Brown, Jaime Ibáñez, Elias Casula, Vishal Rawji, Vincenzo Di Lazzaro, Giacomo Koch, and John Rothwell. 2021. "Disentangling EEG Responses to TMS Due to Cortical and Peripheral Activations." *Brain Stimulation* 14(1):4–18.

Rogasch, Nigel C., Caley Sullivan, Richard H. Thomson, Nathan S. Rose, Neil W. Bailey, Paul B. Fitzgerald, Faranak Farzan, and Julio C. Hernandez-Pavon. 2017. "Analysing Concurrent Transcranial Magnetic Stimulation and Electroencephalographic Data: A Review and Introduction to the Open-Source TESA Software." *NeuroImage* 147:934–51.

Rosanova, Mario, Adenauer Casali, Valentina Bellina, Federico Resta, Maurizio Mariotti, and Marcello Massimini. 2009. "Natural Frequencies of Human Corticothalamic Circuits." *The Journal of Neuroscience: The Official Journal of the Society for Neuroscience* 29(24):7679–85.

Rosanova, Mario, Olivia Gosseries, Silvia Casarotto, Mé Lanie Boly, Adenauer G. Casali, Marie-Auré Lie Bruno, Maurizio Mariotti, Pierre Boveroux, Giulio Tononi, Steven Laureys, and Marcello Massimini. 2012. "Recovery of Cortical Effective Connectivity and Recovery of Consciousness in Vegetative Patients." *Brain* 135:1308–20.

Rossini, P. M., D. Burke, R. Chen, L. G. Cohen, Z. Daskalakis, R. Di Iorio, V. Di Lazzaro, F. Ferreri, P. B. Fitzgerald, M. S. George, M. Hallett, J. P. Lefaucheur, B. Langguth, H. Matsumoto, C. Miniussi, M. A. Nitsche, A. Pascual-Leone, W. Paulus, S. Rossi, J. C. Rothwell, H. R. Siebner, Y. Ugawa, V. Walsh, and U. Ziemann. 2015. "Non-Invasive Electrical and Magnetic Stimulation of the Brain, Spinal Cord, Roots and Peripheral

Nerves: Basic Principles and Procedures for Routine Clinical and Research Application: An Updated Report from an I.F.C.N. Committee." *Clinical Neurophysiology* 126(6):1071–1107.

Salo, Karita S. T., Tuomas P. Mutanen, Selja M. I. Vaalto, and Risto J. Ilmoniemi. 2020. "EEG Artifact Removal in TMS Studies of Cortical Speech Areas." *Brain Topography* 33(1):1–9.

Sarasso, Simone, Mario Rosanova, Adenauer G. Casali, Silvia Casarotto, Matteo Fecchio, Melanie Boly, Olivia Gosseries, Giulio Tononi, Steven Laureys, and Marcello Massimini. 2014. "Quantifying Cortical EEG Responses to TMS in (Un)Consciousness." *Clinical EEG and Neuroscience* 45(1):40–49.

Schambra, Heidi M., R. Todd Ogden, Isis E. Martínez-Hernández, Xuejing Lin, Y. Brenda Chang, Asif Rahman, Dylan J. Edwards, and John W. Krakauer. 2015. "The Reliability of Repeated TMS Measures in Older Adults and in Patients with Subacute and Chronic Stroke." *Frontiers in Cellular Neuroscience* 9:335. doi: 10.3389/fncel.2015.00335. .

Shrout, Patrick E. 1998. "Measurement Reliability and Agreement in Psychiatry." *Statistical Methods in Medical Research* 7(3):301–17.

Shrout, Patrick E., and Joseph L. Fleiss. 1979. "Intraclass Correlations: Uses in Assessing Rater Reliability." *Psychological Bulletin* 86(2):420–28.

Siebner, Hartwig R., Til O. Bergmann, Sven Bestmann, Marcello Massimini, Heidi Johansen-Berg, Hitoshi Mochizuki, Daryl E. Bohning, Erie D. Boorman, Sergiu Groppa, Carlo Miniussi, Alvaro Pascual-Leone, Reto Huber, Paul C. J. Taylor, Risto J. Ilmoniemi, Luigi De Gennaro, Antonio P. Strafella, Seppo Kähkönen, Stefan Klöppel, Giovanni B. Frisoni, Mark S. George, Mark Hallett, Stephan A. Brandt, Matthew F. Rushworth, Ulf Ziemann, John C. Rothwell, Nick Ward, Leonardo G. Cohen, Jürgen Baudewig, Tomáš Paus, Yoshikazu Ugawa, and Paolo M. Rossini. 2009. "Consensus Paper: Combining Transcranial Stimulation with Neuroimaging." *Brain Stimulation* 2(2):58–80.

Streiner, D. L., and G. R. Norman. 2016. "Health Measurement Scales: A Practical Guide to Their Development and Use (5th Edition)." *Australian and New Zealand Journal of Public Health* 40(3):294–95.

Terwee, Caroline B., Sandra D. M. Bot, Michael R. de Boer, Daniëlle A. W. M. van der Windt, Dirk L. Knol, Joost Dekker, Lex M. Bouter, and Henrica C. W. de Vet. 2007. "Quality Criteria Were Proposed for Measurement Properties of Health Status Questionnaires." *Journal of Clinical Epidemiology* 60(1):34–42.

Tiitinen, Hannu, Juha Virtanen, Risto J. Ilmoniemi, Janne Kamppuri, Marko Ollikainen, Jarmo Ruohonen, and Risto Näätänen. 1999. "Separation of Contamination Caused by Coil Clicks from Responses Elicited by Transcranial Magnetic Stimulation." *Clinical Neurophysiology* 110(5):982–85.

Tremblay, Sara, Nigel C. Rogasch, Isabella Premoli, Daniel M. Blumberger, Silvia Casarotto, Robert Chen, Vincenzo Di Lazzaro, Faranak Farzan, Fabio Ferrarelli, Paul B. Fitzgerald, Jeanette Hui, Risto J. Ilmoniemi, Vasilios K. Kimiskidis, Dimitris Kugiumtzis, Pantelis Lioumis, Alvaro Pascual-Leone, Maria Concetta Pellicciari, Tarek Rajji, Gregor Thut, Reza Zomorrodi, Ulf Ziemann, and Zafiris J. Daskalakis. 2019. "Clinical Utility and Prospective of TMS–EEG." *Clinical Neurophysiology* 130(5):802–44.

de Vet, Henrica C. W., Caroline B. Terwee, Dirk L. Knol, and Lex M. Bouter. 2006. "When to Use Agreement versus Reliability Measures." *Journal of Clinical Epidemiology* 59(10):1033–39.

Weir, Joseph P. 2005. "Quantifying Test-Retest Reliability Using the Intraclass Correlation Coefficient and the SEM." *Journal of Strength and Conditioning Research* 19(1):231–40.

Wolf, Steven L., Andrew J. Butler, Georgette I. Campana, Trinity A. Parris, Danielle M. Struys, Sarah R. Weinstein, and Paul Weiss. 2004. "Intra-Subject Reliability of Parameters Contributing to Maps Generated by Transcranial Magnetic Stimulation in Able-Bodied Adults." *Clinical Neurophysiology* 115(8):1740–47.

Wu, Wei, Corey J. Keller, Nigel C. Rogasch, Parker Longwell, Emmanuel Shpigel, Camarin E. Rolle, and Amit Etkin. 2018. "ARTIST: A Fully Automated Artifact Rejection Algorithm for Single-Pulse TMS-EEG Data." *Human Brain Mapping* 39(4):1607–25.

Ye, Song, Keiichi Kitajo, and Katsunori Kitano. 2019. "Information-Theoretic Approach to Detect Directional Information Flow in EEG Signals Induced by TMS." *Neuroscience Research*.

Zazio, Agnese, Guido Barchiesi, Clarissa Ferrari, Eleonora Marcantoni, and Marta Bortoletto. 2022. "M1-P15 as a Cortical Marker for Transcallosal Inhibition: A Preregistered TMS-EEG Study." *Frontiers in Human Neuroscience* 16.

# Tables

**Table 1: Literature review on TMS–EEG measures' reliability.**

| Publication | N | Target | Interval between tests | TMS-EEG measure | Reliability test |
|---|---|---|---|---|---|
| Lioumis et al. 2009 | 7 | M1 (APB) - DLPFC | 1 week | TEPs peaks amplitude and latency | T tests and Pearson's R |
| Casarotto et al. 2010 | 10 | Left occipital, parietal and frontal lobes | same day or 1 week | TEPs | DI and ROC |
| Farzan et al. 2010 | 36 | M1 (APB) - DLPFC | 1 week | LICI | Cronbach's alpha |
| Kerwin et al. 2018 | 16 | DLPFC | 2 recordings 5 minutes apart in the first session. Repeated 1 week after | TEPs peaks amplitude and latency | CCC – SDC |
| ter Braack et al. 2019 | 18 | r-l M1 | 5 recordings in a day to assess effect of daytime on TEP. Same protocol repeated 1 week later for 3 subjs | Continuous TEPs and TEPs amplitude/latency | T tests (cluster-based) |
| Ye et al. 2019 | 12 | Oz | | VAR and STE | Spearman's R |
| Casula et al. 2021 | 50 | r-l M1 (FDI) | 3 weeks | IHP and ISP | ICC |
| Ozdemir et al. 2020 | 21 | DMN (r-AG) and DAN (r-SP) | 4 weeks | TEPs | Pearson's R |
| de Goede et al. 2020 | 25 | r-l M1 (ADM) | 1 week | Continuous TEPs and TEPs amplitude/latency | T tests/ICC (3,1) |
| Momi et al. 2021 | 21 | DMN (r-AG) and DAN (r-SP) | 4 weeks | TEPs | Pearson's R |
| Mancuso et al. 2021 | 8 | l-M1 - r-mPFC | 1 week | TEPs | T tests – CCC |
| Ozdemir et al. 2021 | 24 | l-IPL, l-M1 (FDI) and l-DLPFC | 1 month | TEPs | SI |
| Ozdemir et al. 2021 | 10 | l-IPL and l-M1 (FDI) (l-DLPFC) | 2 months (2.5 weeks) | TEPs | SI |
| Bertazzoli et al. 2021 | 16 | l-DLFPC - l-IPL | 72.3 ±35.8 days | TEPs | CCC - ANOVA - Spearman's R |

*r = right; l = left; M1 = primary motor cortex; APB = abductor pollicis brevis; DLPFC = dorsolateral prefrontal cortex; FDI = first dorsal interosseus; DMN = default-mode network; DAN = dorsal attention network; mPFC = medial prefrontal cortex; IPL = inferior parietal lobule; LICI = long-interval cortical inhibition; VAR = vector autoregression; STE = symbolic transfer of entropy; IHP = inter-hemispheric balance; ISP = inter-hemispheric propagation; DI = divergence index; ROC = receiver operator characteristic; CCC = concordance correlation coefficient; SI = similarity index.*

**Table 2: Fictional data for demonstrating the relation between ICC and SEMeas/SDC.**

**A**

Example Data

| | T0 | T1 | diff |
|---|---|---|---|
| s1 | 25 | 27 | 2 |
| s2 | 20 | 30 | 10 |
| s3 | 7 | 10 | 3 |
| s4 | 40 | 35 | -5 |
| s5 | 45 | 40 | -5 |
| s6 | 30 | 35 | 5 |
| s7 | 15 | 25 | 10 |
| s8 | 20 | 15 | -5 |
| Mean | 25.25 | 27.12 | |
| Stnd. Dev | 11.85 | 9.63 | |

ANOVA

| Source of variance | SS | df | MS | F | sign. | F crit |
|---|---|---|---|---|---|---|
| Between subjects | 1723.93 | 7 | 246.27 | 12.10 | 0.00196 | 3.78 |
| Within subjects | 156.5 | 8 | 19.56 | | | |
| Time | 14.06 | 1 | 14.06 | 0.69 | 0.433 | 5.59 |
| Error | 142.43 | 7 | 20.34 | | | |
| Total | 1880.43 | 15 | | | | |
| ICC | 0.85 | | | | | |
| SEMeas | 4.42 | | SDC | | 12.258 | |

**B**

Example Data

| | T0 | T1 | diff |
|---|---|---|---|
| S1 | 25 | 26 | 1 |
| S2 | 20 | 22 | 2 |
| S3 | 7 | 10 | 3 |
| S4 | 40 | 38 | -2 |
| S5 | 45 | 43 | -2 |
| S6 | 30 | 30 | 0 |
| S7 | 15 | 16 | 1 |
| s8 | 20 | 22 | 2 |
| Mean | 25.25 | 25.87 | |
| Stnd. Dev | 11.85 | 10.22 | |

ANOVA

| Source of variance | SS | df | MS | F | sign. | F crit |
|---|---|---|---|---|---|---|
| Between subjects | 1948.43 | 7 | 278.34 | 163.21 | 3.24E-07 | 3.78 |
| Within subjects | 13.5 | 8 | 1.68 | | | |
| Time | 1.56 | 1 | 1.56 | 0.91 | 0.37 | 5.59 |
| Error | 11.95 | 7 | 1.70 | | | |
| Total | 1961.93 | 15 | | | | |
| ICC | 0.98 | | | | | |
| SEMeas | 1.2990 | | SDC | | 3.60 | |

**C**

Example Data

| | T0 | T1 | diff |
|---|---|---|---|
| s1 | 25 | 27 | 2 |
| s2 | 23 | 33 | 10 |
| s3 | 19 | 22 | 3 |
| s4 | 33 | 28 | -5 |
| s5 | 35 | 30 | -5 |
| s6 | 20 | 25 | 5 |
| s7 | 17 | 27 | 10 |
| s8 | 24 | 19 | -5 |
| Mean | 24.5 | 26.37 | |
| Stnd. Dev | 6.04 | 4.12 | |

ANOVA

| Source of variance | SS | df | MS | F | sign. | F crit |
|---|---|---|---|---|---|---|
| Between subjects | 285.43 | 7 | 40.77 | 2.003 | 0.1896 | 3.78 |
| Within subjects | 156.5 | 8 | 19.56 | | | |
| Time | 14.06 | 1 | 14.06 | 0.69 | 0.433 | 5.59 |
| Error | 142.43 | 7 | 20.34 | | | |
| Total | 441.93 | 15 | | | | |
| ICC | 0.35 | | | | | |
| SEMeas | 4.42 | | SDC | | 12.259 | |

**D**

Example Data

| | T0 | T1 | diff |
|---|---|---|---|
| s1 | 25 | 26 | 1 |
| s2 | 24 | 27 | 3 |
| s3 | 23 | 26 | 3 |
| s4 | 26 | 24 | -2 |
| s5 | 25 | 23 | -2 |
| s6 | 23 | 23 | 0 |
| s7 | 22 | 23 | 1 |
| s8 | 27 | 29 | 2 |
| Mean | 24.37 | 25.12 | |
| Stnd. Dev | 1.57 | 2.08 | |

ANOVA

| Source of variance | SS | df | MS | F | sign. | F crit |
|---|---|---|---|---|---|---|
| Between subjects | 41 | 7 | 5.85 | 2.98 | 0.086 | 3.78 |
| Within subjects | 16 | 8 | 2 | | | |
| Time | 2.25 | 1 | 2.25 | 1.14 | 0.32 | 5.59 |
| Error | 13.75 | 7 | 1.96 | | | |
| Total | 57 | 15 | | | | |
| ICC | 0.49 | | | | | |
| SEMeas | 1.41 | | SDC | | 3.92 | |

**Table 3: Fictional data showing the relation between ICC, SEMeas/SDC and Pearson's**

| HIGH Pearson's R - LOW ICC | | | | | HIGH Pearson's R - LOW ICC | | | |
|---|---|---|---|---|---|---|---|---|
| | T0 | T1 | Diff | | | T0 | T1 | diff |
| s1 | 25 | 29 | 4 | | S1 | 25 | 15 | -10 |
| S2 | 24 | 25 | 1 | | S2 | 30 | 19 | -11 |
| S3 | 27 | 31 | 4 | | S3 | 41 | 28 | -13 |
| S4 | 25 | 27 | 2 | | s4 | 23 | 15 | -8 |
| S5 | 28 | 33 | 5 | | s5 | 28 | 16 | -12 |
| S6 | 30 | 37 | 7 | | s6 | 30 | 24 | -6 |
| S7 | 28 | 33 | 5 | | s7 | 28 | 11 | -17 |
| s8 | 27 | 30 | 3 | | s8 | 20 | 10 | -10 |
| Mean | 26.75 | 30.62 | | | Mean | 28.12 | 17.25 | |
| Stnd. Dev | 1.85 | 3.53 | | | Stnd. Dev | 5.86 | 5.78 | |

**ANOVA**

| Source of variance | SS | df | MS | F | sign. | F crit |
|---|---|---|---|---|---|---|
| Between subjects | 114.93 | 7 | 16.41 | 9.24 | 0.00445 | 3.78 |
| Within subjects | 72.5 | 8 | 9.06 | | | |
| Time | 60.06 | 1 | 60.06 | 33.80 | 0.00065 | 5.591 |
| Error | 12.43 | 7 | 1.77 | | | |
| Total | 187.43 | 15 | | | | |
| ICC | 0.28 | | | | | |
| SEMeas | 3.01 | | SDC | 8.34 | | |
| Pearson's R | 0.97 | | | | | |

**ANOVA**

| Source of variance | SS | df | MS | F | sign. | F crit |
|---|---|---|---|---|---|---|
| Between subjects | 503.93 | 7 | 71.99 | 13.11 | 0.001532 | 3.78 |
| Within subjects | 511.5 | 8 | 63.93 | | | |
| Time | 473.06 | 1 | 473.06 | 86.15 | 3.49E-05 | 5.59 |
| Error | 38.43 | 7 | 5.49 | | | |
| Total | 1015.43 | 15 | | | | |
| ICC | 0.05 | | | | | |
| SEMeas | 7.99 | | SDC | 22.16 | | |
| Pearson's R | 0.85 | | | | | |

**Figures and Figures' Captions**

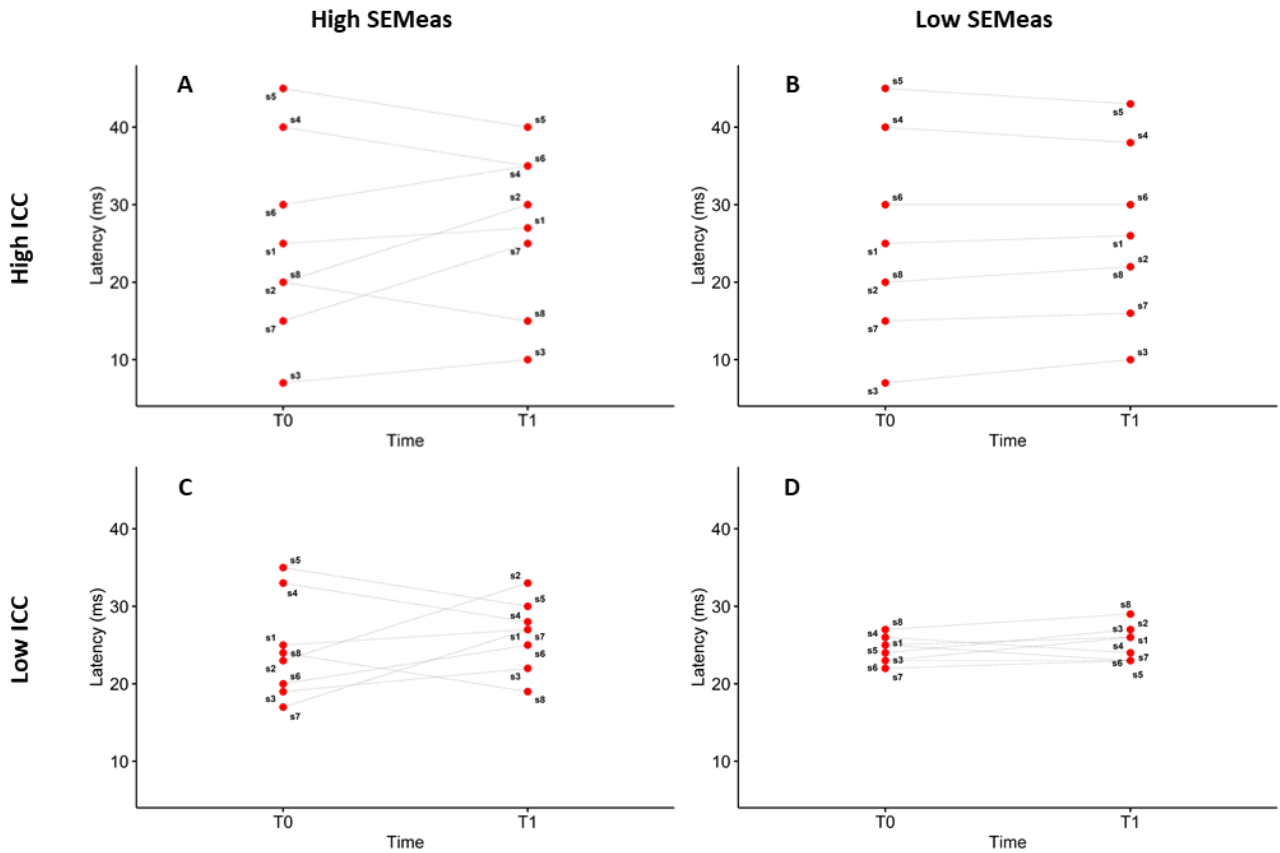

**Figure 1:** Graphical representation of the test-retest fictional data presented in Table 2. The Y-axis represents the latency (ms) of a TEP component. The X-axis represents the test (T0) and retest (T1) sessions.

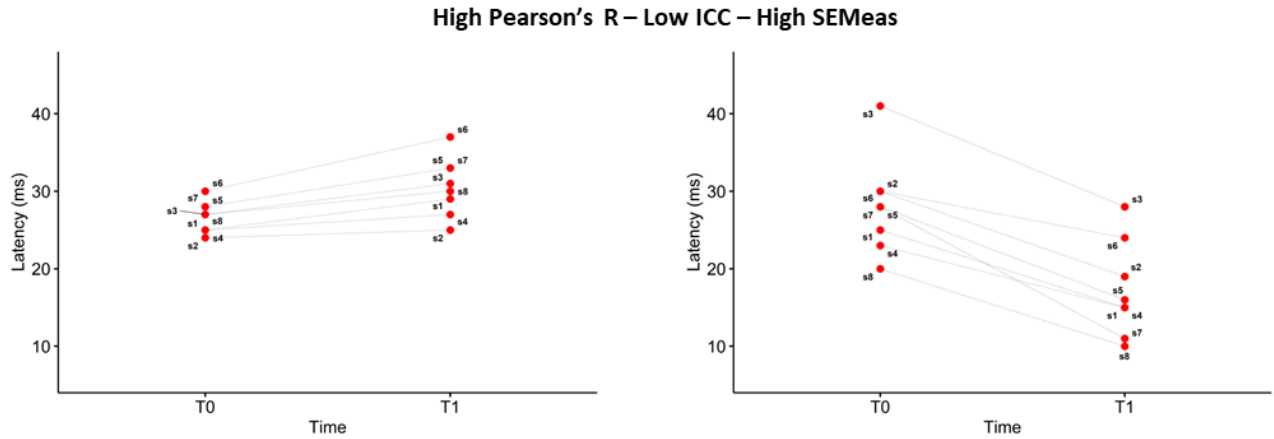

**Figure 2:** Graphical representation of the test-retest fictional data presented in Table 3. The Y-axis represents the latency (ms) of a TEP component. The X-axis represents the test (T0) and retest (T1) sessions. From Table 3, the left panel corresponds to the left graph, and the right panel corresponds to the right graph

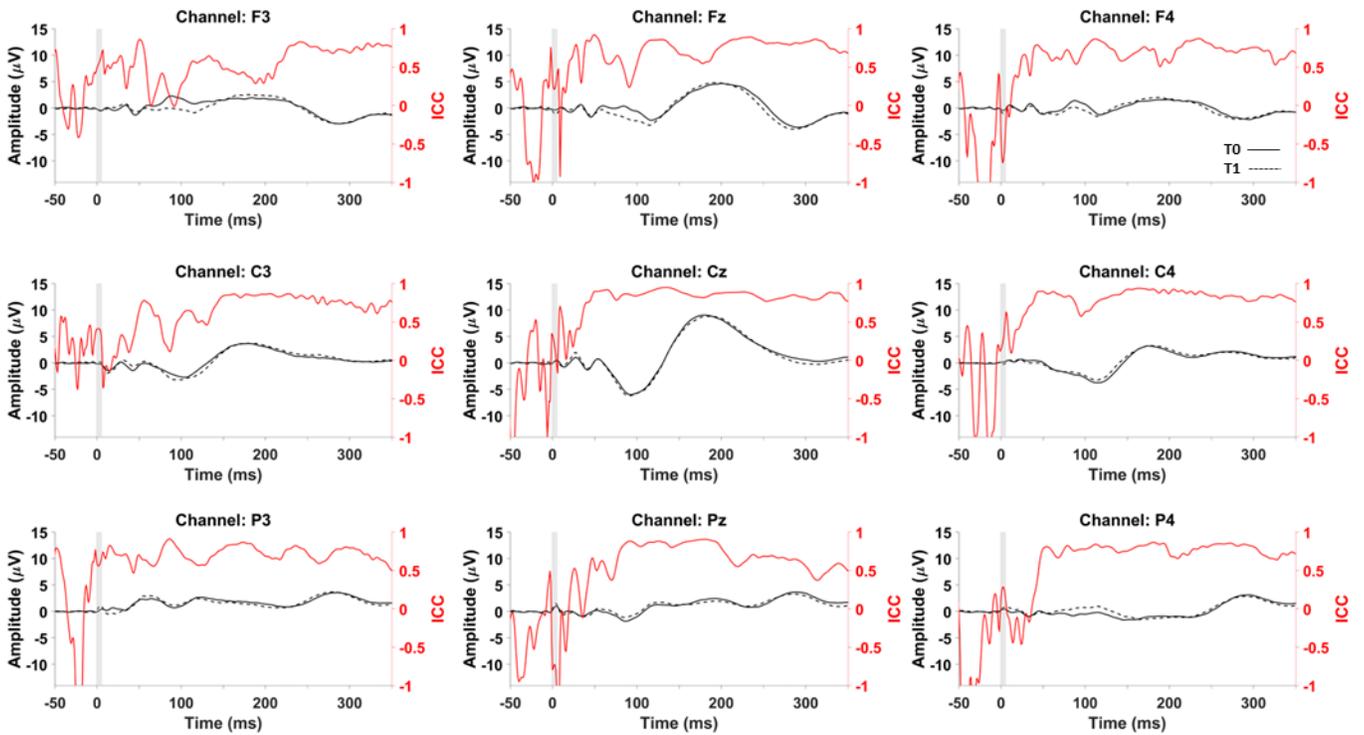

**Figure 3:** ICC over time in 9 selected electrodes. The black line depicts the average across subjects TEP for that electrode in session one, while the black dotted line depicts the same electrode in session 2. The red solid line pictures the ICC calculated at each time point between the two sessions. The red dotted line represents the 0.6 limit for moderate to high reliability (Shrout et al. 1998). TMS was delivered to the left IPL.

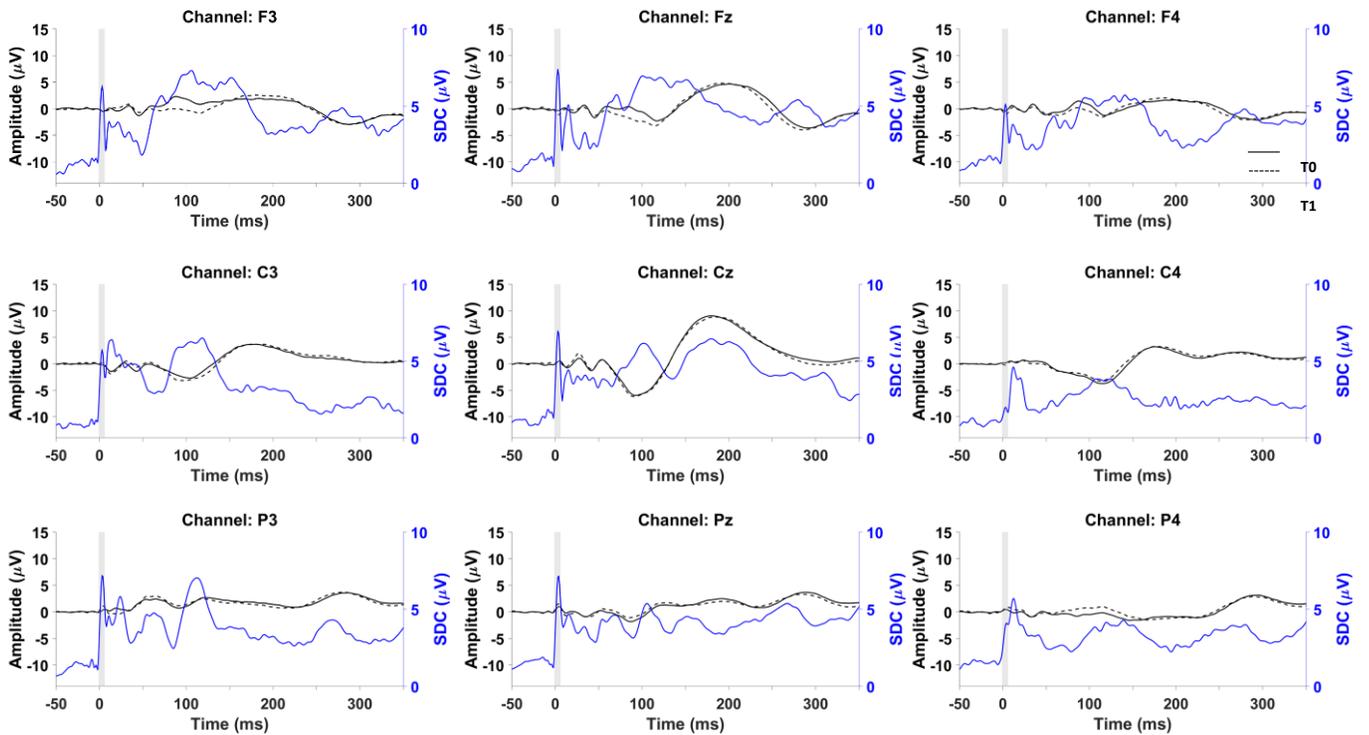

**Figure 4:** SDC over time in 9 selected electrodes. The black line depicts the average across subjects TEP for that electrode in session one, while the black dotted line depicts the same electrode in session 2. The blue solid line pictures the SDC calculated at each time point between the two sessions. TMS was delivered to the left IPL.